\documentclass[journal]{IEEEtran}
\usepackage{multirow}
\usepackage{graphicx}
\usepackage[cmex10]{amsmath}

\begin{document}

\title{Pixelated Geiger-Mode Avalanche Photo-Diode Characterization through Dark Current Measurement}
\author{Pierre-Andr\'e~Amaudruz~\IEEEmembership{Member,~IEEE,}
        Daryl~Bishop,
        Colleen~Gilhully,
        Andrew~Goertzen~\IEEEmembership{Member,~IEEE,} 
        Lloyd~James,
        Piotr Kozlowski, 
        Fabrice~Reti\`ere~\IEEEmembership{Member,~IEEE,}
        Ehsan Shams,
        Vesna Sossi~\IEEEmembership{Member,~IEEE,} 
        Greg Stortz, 
        Jonathan D. Thiessen,
        Christopher J. Thompson ~\IEEEmembership{Life Senior Member, ~IEEE} 
\thanks{P-A Amaudruz, D. Bishop, C. Gilhully, L. James, and F. Reti\`ere are with TRIUMF, Vancouver, British Columbia, Canada, email: fretiere@triumf.ca}%
\thanks{A. Goertzen, E. Shams, and J. D. Thiessen are with the Department of Radiology, University of Manitoba, Winnipeg, Manitoba, Canada}
\thanks{P. Kozlowski is with the Department of Radiology, University of British Columbia, Vancouver, British Columbia, Canada.}
\thanks{V. Sossi and G. Stortz are with the Department of Physics and Astronomy, University of British Columbia, Vancouver, British Columbia, Canada.}
\thanks{C.J. Thompson is with the McConnell Brain Imaging Centre, Montreal Neurological Institute, Montréal, Québec, Canada.}}%

\maketitle

\begin{abstract}

PIXELATED geiger-mode avalanche photodiodes(PPDs), often called silicon photomultipliers (SiPMs) are emerging as an excellent replacement for traditional photomultiplier tubes (PMTs) in a variety of detectors, especially those for subatomic physics experiments, which requires extensive test and operation procedures in order to achieve uniform responses from all the devices. In this paper, we show for two PPD brands, Hamamatsu MPPC and SensL SPM, that the dark noise rate, breakdown voltage and rate of correlated avalanches can be inferred from the sole measure of dark current as a function of operating voltage, hence greatly simplifying the characterization procedure. We introduce a custom electronics system that allows measurement for many devices concurrently, hence allowing rapid testing and monitoring of many devices at low cost. Finally, we show that the dark current of Hamamastu Multi-Pixel Photon Counter (MPPC) is rather independent of temperature at constant operating voltage, hence the current measure cannot be used to probe temperature variations. On the other hand, the MPPC current can be used to monitor light source conditions in DC mode without requiring strong temperature stability, as long as the integrated source brightness is comparable to the dark noise rate.

\end{abstract}


\IEEEpeerreviewmaketitle

\section{Introduction}

\IEEEPARstart{P}{ixelated} geiger-mode avalanche photodiodes (PPDs), often called silicon photomultipliers (SiPMs) are emerging as an excellent replacement for traditional photomultiplier tubes (PMTs) in a variety of detectors, especially those for subatomic physics experiments \cite{GaruttiSiPMHEP} and medical imaging \cite{SiPMNuclMed}. The T2K near detector experiment was one of the first to take advantage of them, using around 50,000 Multi-Pixel Photon Counters (MPPCs), the Hamamatsu Photonics brand name for PPDs \cite{T2KExp}. Because T2K was an early adopter, numerous control and calibration tools were implemented to achieve stable operation. The Fine Grained Detector for example, relies on more than 1,000 temperature sensors to ensure stable operation of 8,448 MPPCs \cite{FGD}. The electronics were also set up to allow continuous monitoring of the MPPC gain by detecting dark noise generated pulses that have the same charge as those caused by single photo-electrons \cite{FGD}. Gross properties of all MPPCs were measured prior to installation \cite{mppcQA} and a few MPPCs were characterized in detail to assess their possible contributions to physics observables (which turned out to be very small) \cite{T2Kmppc}. Information on the principles of operation of PPDs may be found in other papers \cite{Bondarenko}, \cite{Dolgoshein}, \cite{Golovin}, \cite{RenkerA}, \cite{Renker}, \cite{Sadygov}. In this paper, we investigate using only the measure of current as a function of voltage for monitoring and characterizing MPPCs, hence greatly simplifying their operation and calibration. We also investigate whether or not the MPPC temperature can be measured by recording the current at fixed operating voltage, which would alleviate the need for independent temperature sensors. To further support the conclusions of this paper, data collected using a similar process with SensL SPMs \cite{SensL} are also presented.

This work was motivated by the needs of experiments that followed the construction of the T2K FGD electronics system that included extensive monitoring and control capabilities. The T2K electronics were first reconfigured to equip a prototype detector for positron emission tomography\cite{WavePET}. It was then expanded to the prototype target system for the TREK experiment foreseen at J-PARC \cite{TREK}. Then, new boards were designed to equip the first spectrometer for Muon Spin Rotation experiments to use PPDs at TRIUMF following a similar design at Paul Sherrer Institute \cite{PSI}. The slow control electronics system has proven well suited to operate MPPCs, whether using 32 as in the muSR spectrometer, or thousands as in T2K. In this paper we will first describe the electronics system and experimental setup that was used to take data. Then, in section 3, we will show that detailed information about MPPCs can be obtained from current-voltage measurement. In section 4, we will investigate using the MPPCs themselves as temperature sensors and we will conclude in section 5.

\section{Experimental Setup}
\subsection{Electronics}
\subsubsection{Hamamatsu MPPCs}
In this investigation, 8 1x4 mm$^2$ MPPCs (model S10984-050P) and 8 3x3 mm$^2$ MPPCs (model S10931-050P) were characterized. These MPPCs were attached to plastic mounts. An 8 channel custom printed circuit board designed for a muon spin rotation ($\mu SR$ ) spectrometer was used to supply high voltage to the MPPCs, control the bias voltage, and gather current measurements. The ability to set individual bias voltage offsets through the Digital-Analog Converter (16-bit Linear Technologies LTC2605 DAC) was not used in the context of this paper. Communications with the microcontroller on board were done via a MIDAS Slow Control Bus (MSCB) \cite{mscb}. Access to the control card was possible via command line or through C programs using a library of MSC functions.  

\begin{figure}[t]
  \centering
 \includegraphics[width=0.6\textwidth, angle=0]{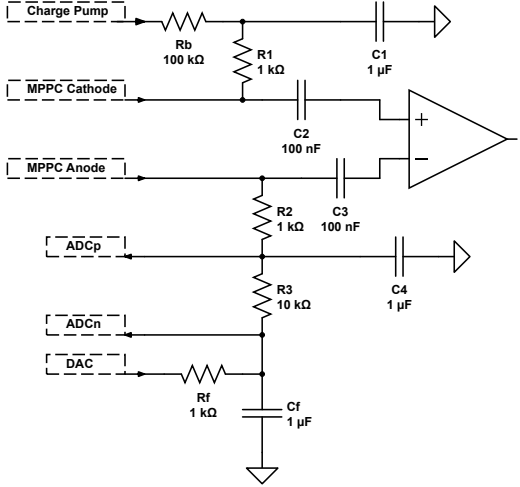}
  \caption{Diagram of the circuitry involved in measuring the bias voltage and the current from the MPPC.}
  \label{fig:circuit}
\end{figure}

As can be seen in Figure~\ref{fig:circuit}, a charge pump is used to generate a bias voltage of up to 74 V. The charge pump voltage is then fed to all 8 channels on board. A 100k$\Omega$ resistor in series with each MPPC allows continued operation even if a short circuit occurs in one of the 8 channels. The 1k$\Omega$ resistors R1 and R2 are used to ensure that the current from the fast pulses flows through the high speed amplifier (LTC6401-26) that has a 50$\Omega$ input impedance rather than through the bias lines. The current is inferred by measuring the voltage drop across a high precision 10 k$\Omega$ resistor using a  8-channel, 24-bit ADC (Linear Technologies LTC2499). The multiplexer out (MUXOUT) lines on the ADC were connected to a high impedance Linear Technologies LTC6078CMS8 amplifier, and then this was connected back to the ADC input (ADCIN) lines on the ADC. This high impedance connection ensures that the current flowing through the ADC is negligible compared to the current being measured.

\subsubsection{SensL PPDs}
In addition to Hamamatsu MPPCs, the MPPC characterization methods were tested on two generations of SensL 4x4 PPD arrays (SPMArray4 and ArraySL-4, SensL, Cork, Ireland).  Both arrays consist of a 4x4 arrangement of Geiger-mode avalanche photodiodes with a typical operating voltage of ~29.5 V (+2 V above the breakdown voltage), with small differences in dimensions and gain between the first generation (SPMArray4) and second generation (ArraySL-4) devices.  In the SPMArray4, each photodiode has an active area of 2.85 x 2.85 mm$^2$ with 3640 microcells per photodiode (using Hamamatsu’s terminology, “microcells” are equivalent to “pixels”) and a typical gain of 1x$10^6$.  In the ArraySL-4, each photodiode has an active area of 3.05 x 3.05 mm$^2$ with 4774 microcells per photodiode and a typical gain of 2.4x$10^6$.  The photodiodes are mounted on a ceramic package with a total effective area of 13.4 x 13.4 mm$^2$ and a 20-pin grid array for electrical I/O.

All measurements of the SensL 4x4 PPD arrays were done using a front-end detector card which multiplexes the photodiode signals from 16 to 4 with a charge division resistor network and uses an HDMI\textregistered ~connection for power supply and analog signal readout \cite{HDMI}.  Temperature was monitored underneath the PPD socket with a ±$2^\circ$ C precision temperature-to-voltage converter (Microchip Technology TC1047A).  Differential signals and temperature data were read out and bias voltage and $\pm 2.8$ V amplifier power were supplied via a 2m-long HDMI\textregistered ~v1.4 Type A to Type C cable (BrightLink Cables) connected to a receiver card. The receiver card has the ability to set bias voltage via a step-up voltage regulator (Linear Technologies LT3482) which creates a 35V supply voltage with an op-amp (Texas Instruments OPA237NA/250) supplying bias voltage to each detector, controlled by a digital-to-analog converter (DAC) (Linear Technology LTC2619CGN). Bias voltage, stepped down with a voltage divider, and bias current, amplified with a current sense amplifier (Linear Technologies LT6100), were both recorded with an analog-to-digital converter (ADC) (Linear Technologies LTC2499CUHF), as was temperature data from the transmitter card.  The ADC and DAC were controlled and monitored via the I2C bus with a Model B Raspberry Pi running a custom-built slow control GUI written in Python 3.

\subsection{Test Chamber}
\subsubsection{Hamamatsu MPPCs}
A MicroClimate Test Chamber from Cincinnati Sub-Zero with a Watlow F4 controller was used to set a variety of constant temperatures for the MPPCs. The chamber also acted as a dark box. To measure the dark current effectively, the MPPCs must not be exposed to light or ionizing radiation that could induce current avalanches. The MPPCs and plastic mounts were inside the test chamber, connected with micro coaxial cables to the control card outside the chamber. The control card was left outside to avoid introducing a heat source into the chamber. The MPPCs themselves dissipate very little heat. Due to the short length of commercially available cables, the MPPCs were not positioned in the centre of the temperature chamber, but were very close to the cable port. This port is sealed with an insulating foam plug. It is very likely that the temperature at this location was somewhat closer to room temperature than the measurement given by the chamber controller. This uncertainty regarding temperature is the primary systematic error in this investigation.

\subsubsection{SensL PPDs}
For temperature-dependent measurements with the SensL PPDs, the transmitter board was enclosed in a small, light-tight aluminum enclosure (Hammond Manufacturing) with inlets for an HDMI\textregistered ~cable and cold air supply.  The aluminum enclosure was cooled with a vortex tube (Exair Model \# 3202 with a 2 SCFM generator) connected to an air compressor (Snap-On, 20 gallon tank with a 2 HP motor capable of 4.1 CFM@90 PSI).  The air temperature from the cold air outlet of the vortex tube can be adjusted by changing the supply air pressure and the cold air fraction of the vortex tube, allowing us to cool the aluminum enclosure down to ~$6^\circ$ C.  Heating was accomplished with a heating pad (Fisher Scientific Isotemp 11-100-49SH) underneath the aluminum enclosure.  Although this setup didn’t provide the same temperature span and stability as an environmental testing chamber, it emulated the temperature ranges and temperature control capabilities of a pre-clinical positron emission tomography system for which the SensL PPD arrays were tested.

\section{Using I-V Curves to Characterize MPPCs}

The current versus voltage was measured for each PPD across a range of temperatures from 0 $^{\circ}$C to 40 $^{\circ}$C, with a focus on temperatures from 16 $^{\circ}$C to 30 $^{\circ}$C. For the MPPC tests, the bias voltage was scanned by varying the charge pump, using a previously determined linear relationship. The nominal resolution is 20 mV; however, fluctuations in voltage caused some variation in the spacing of data points. The resistors in series with the MPPCs introduce a voltage drop that has to be corrected for in order to estimate the voltage across the photo-diodes ($V_{PPD}$) such as $V_{PPD} = V_{Qpump} - V_{DAC} - R \times  i_{PPD}$  with R=113 k$\Omega$ the sum of resistors in series with the PPDs, and $V_{Qpump}$ and $V_{DAC}$ the voltages provided by the charge pump and the DAC respectively. In the case of the SensL SPMs, the series resistor was only 56$\Omega$ and the voltage drop is negligible.

\begin{figure} [t]
  \centering
\includegraphics[width=0.45\textwidth]{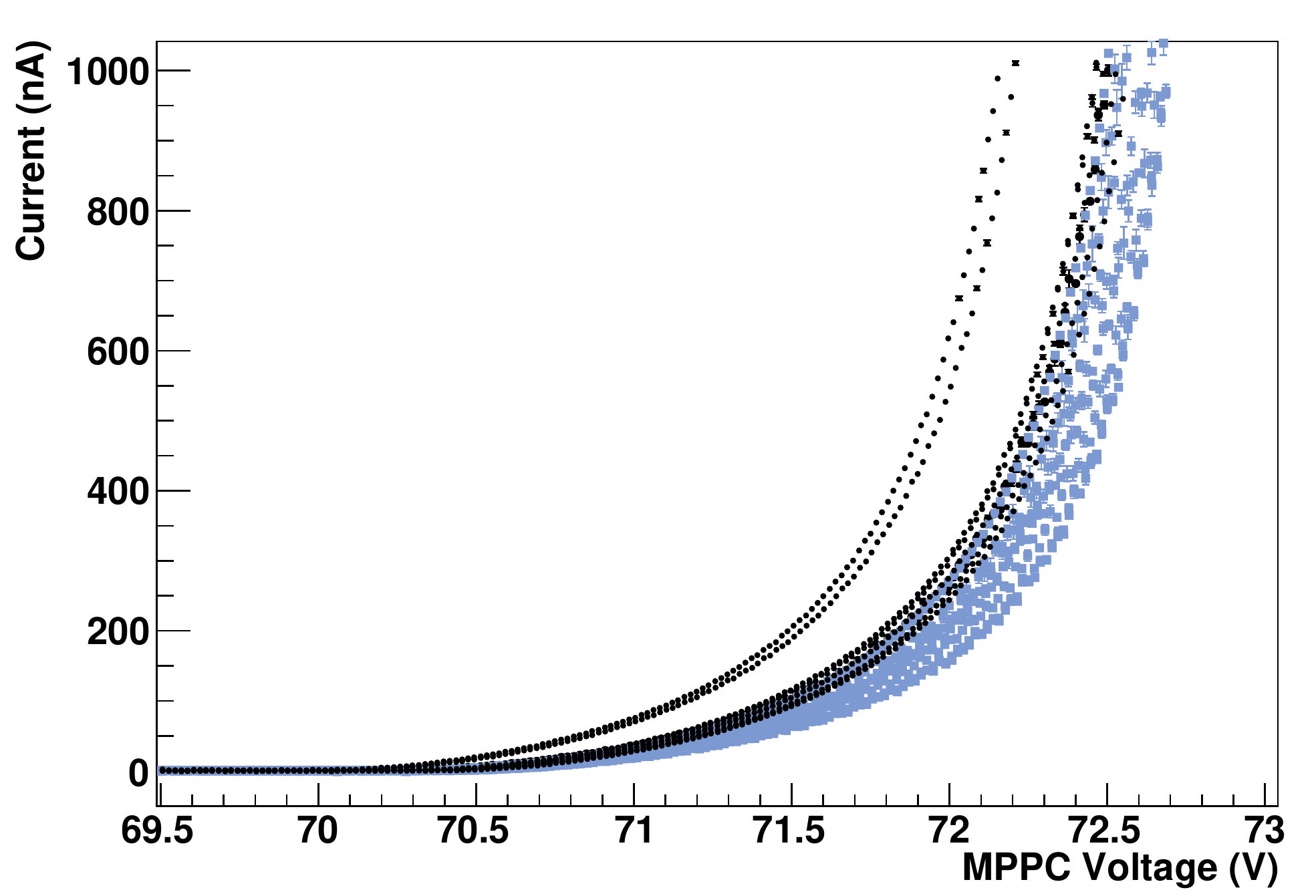}
  \caption{I-V curves for all MPPCs studied at 20 $^\circ$C. The black dots represent the 3x3 mm$^2$ MPPCs, and the blue squares represent the 1x4 mm$^2$ MPPCs.}
  \label{fig:AllMPPCs_T20}
\end{figure}

Figure~\ref{fig:AllMPPCs_T20} shows example I-V curves at 20 $^\circ$C for each of the MPPCs in the absence of any light sources. The current averaged over time is the product of the gain and the dark noise rate, weighted for the probability of correlated avalanches. Expressions for all of these factors were taken from the T2K MPPC characterization paper \cite{T2Kmppc} that investigated various properties of the MPPCs. All the processes were found to depend on  the over-voltage defined as $\Delta V = V_{PPD} - V_{bd}$ with $V_{PPD}$ the voltage across the PPD and $V_{bd}$ the breakdown voltage. The dark noise rate dependence on over-voltage was found to be linear such as $Rate_{DN} = DN(T) \Delta V$ with $DN(T)$ a temperature dependent parameter. In this paper, we neglect the offset found in $V_{bd}$ found in \cite{T2Kmppc} between the linear extrapolation of the gain and dark noise rate. The average number of correlated avalanches caused by one avalanche, as a result of after-pulsing and cross-talk, is defined as $ N_{ca}(T) \cdot \Delta V^2$, where $N_{ca}$(T) is a temperature dependent parameter. Accounting for the possibility of any number $k$ of generations of correlated avalanches (correlated avalanches themselves generating subsquent correlated avalanches), the rate of thermally generated avalanches has to be multiplied by $1/(1-N_{ca} \cdot \Delta V^2)$\cite{Vinogradov}. The avalanche rate is then multiplied by the gain, i.e. the charge per avalanche, to obtain the current. The gain was also found to increase linearly such as $Gain =  C \cdot \Delta V$ with C the pixel capacitance. Below breakdown voltage voltage, the current exhibits an offset that is dependent on the current readout circuit. A fixed constant $\beta$ is used below breakdown voltage. The final fit function has then the following form:
\begin{eqnarray}
I(\Delta V)&=& DN \cdot C \Big(\frac{1}{1-N_{ca}\Delta V^2}\Big)\Delta V^2 \nonumber \\ 
&+& \beta \hspace{5mm} for \Delta V \hspace{1mm} \geq 0; \nonumber \\
&=& \beta \hspace{5mm} for \Delta V \hspace{1mm} < 0
\end{eqnarray}
\begin{figure}[t]
  \centering
 \includegraphics[width=0.45\textwidth, angle=0]{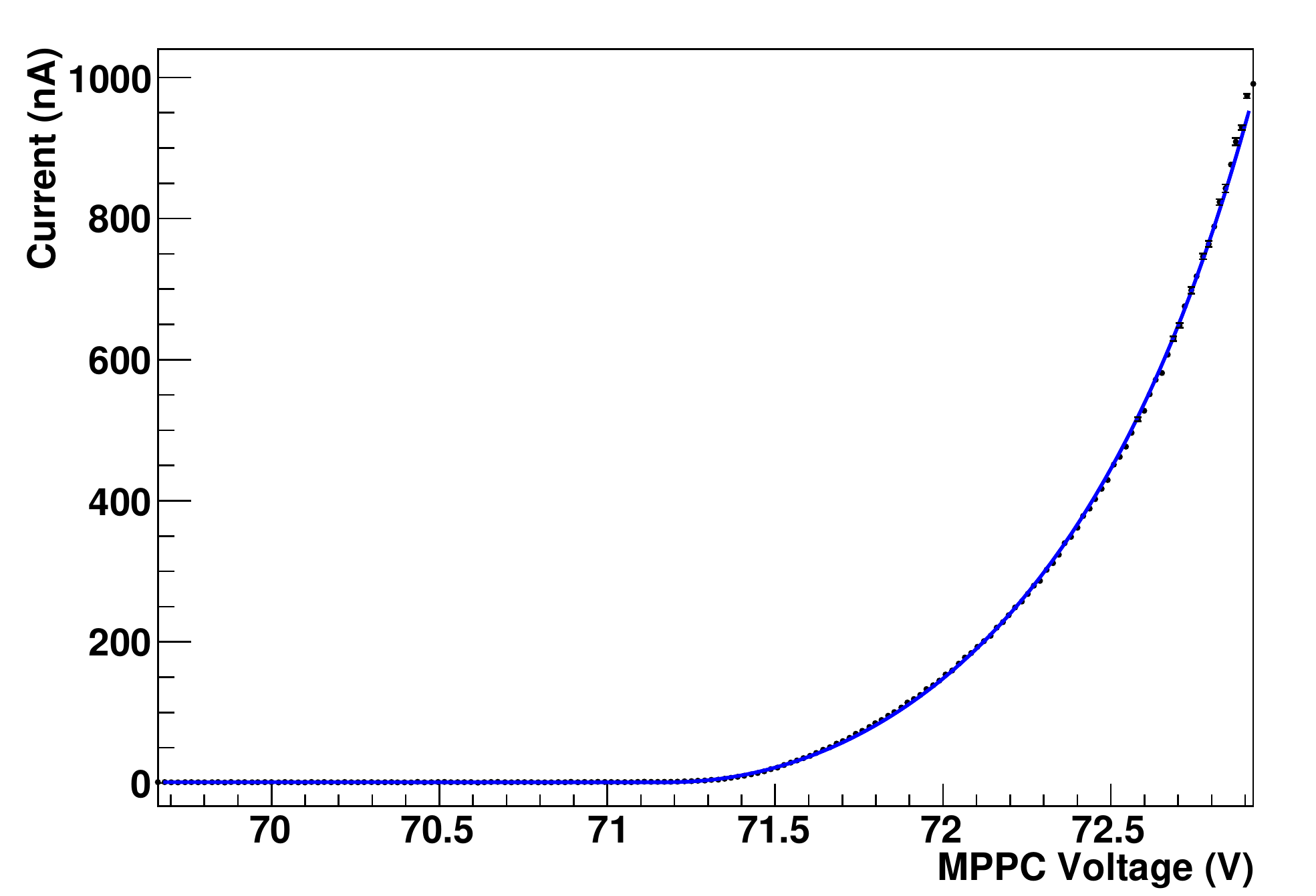}
  \caption{Example of a good quality I-V fit (at 34 $^\circ$C) to 3x3 mm$^2$ MPPC 05.}
  \label{fig:T20_IV}
\vspace{8mm}
\includegraphics[width=0.45\textwidth]{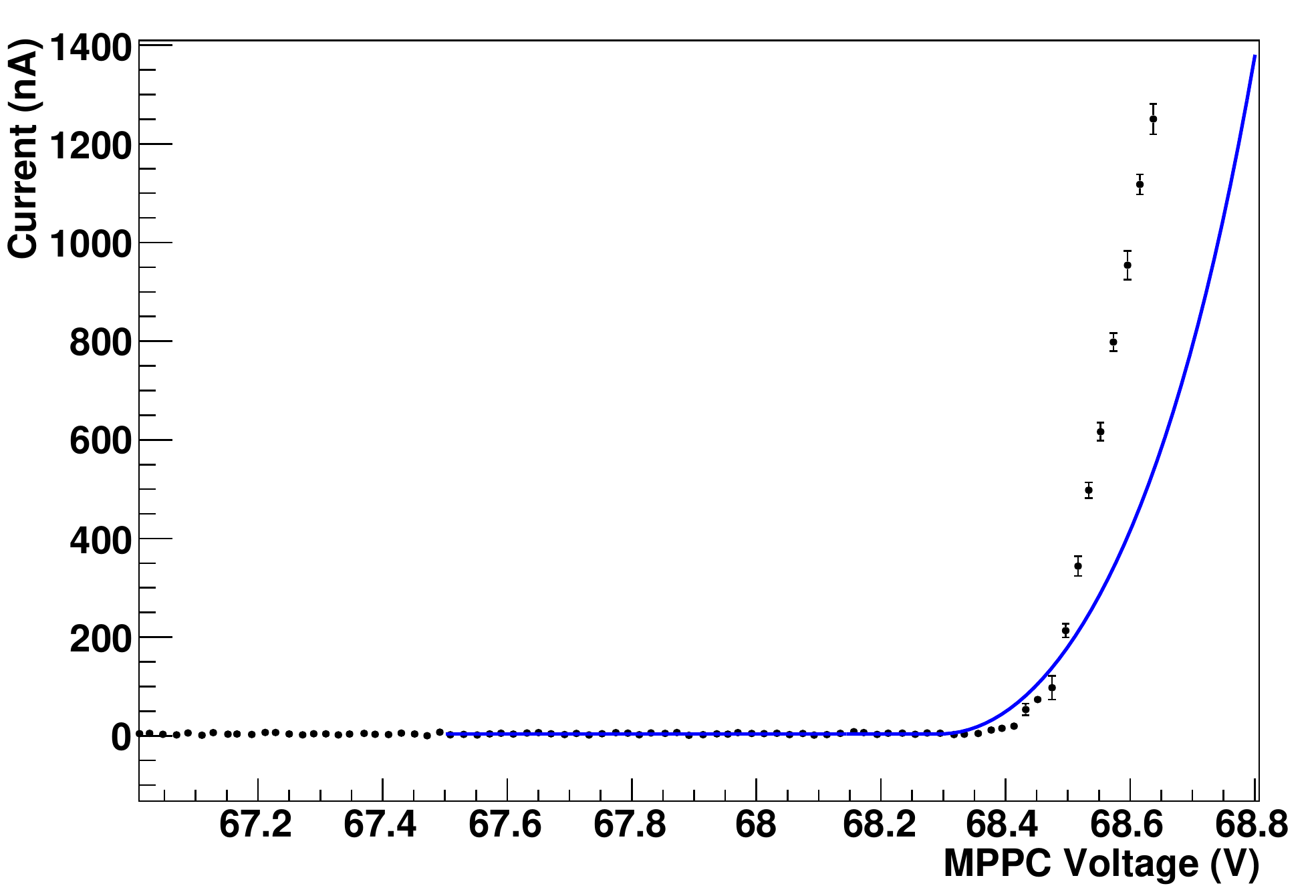}
  \caption{Example of the poor results of fits at low temperatures (-60 $^\circ$C). This is due to the very rapid increase in current after $V_{bd}$ that cannot be described by the expected quadratic expression. }
  \label{fig:T-60_IV}
\end{figure}
An automated fitting routine for I-V curves was created. The first fit extends to the voltage where the current first exceeds 60 nA (for the MPPC data) or 25$\mu$A (for the SensL data). The fit provides an initial estimate of the breakdown voltage. Subsequent fits defined the fitting range based on the previous value of the breakdown voltage, until the value settled or the fitting routine timed out. It was noted that if all data points on an I-V curve were taken into account by the fitting routine, then the resulting fit would not closely match the data between the breakdown voltage and 1.5V above it. This was due to the attempts of the fitting routine to fit the data points at an overvoltage over 2V, where the equation used in the fit matches the data less closely. To account for this, most final fits are based only on points up to 2.25V above $V_{bd}$. $N_{ca}$ was limited between 0 and 1. $DN \cdot C$ was limited to being greater than 0 (with a high upper bound) to prevent meaningless negative values. 

\begin{figure} [t]
  \centering
    \includegraphics[width=0.45\textwidth]{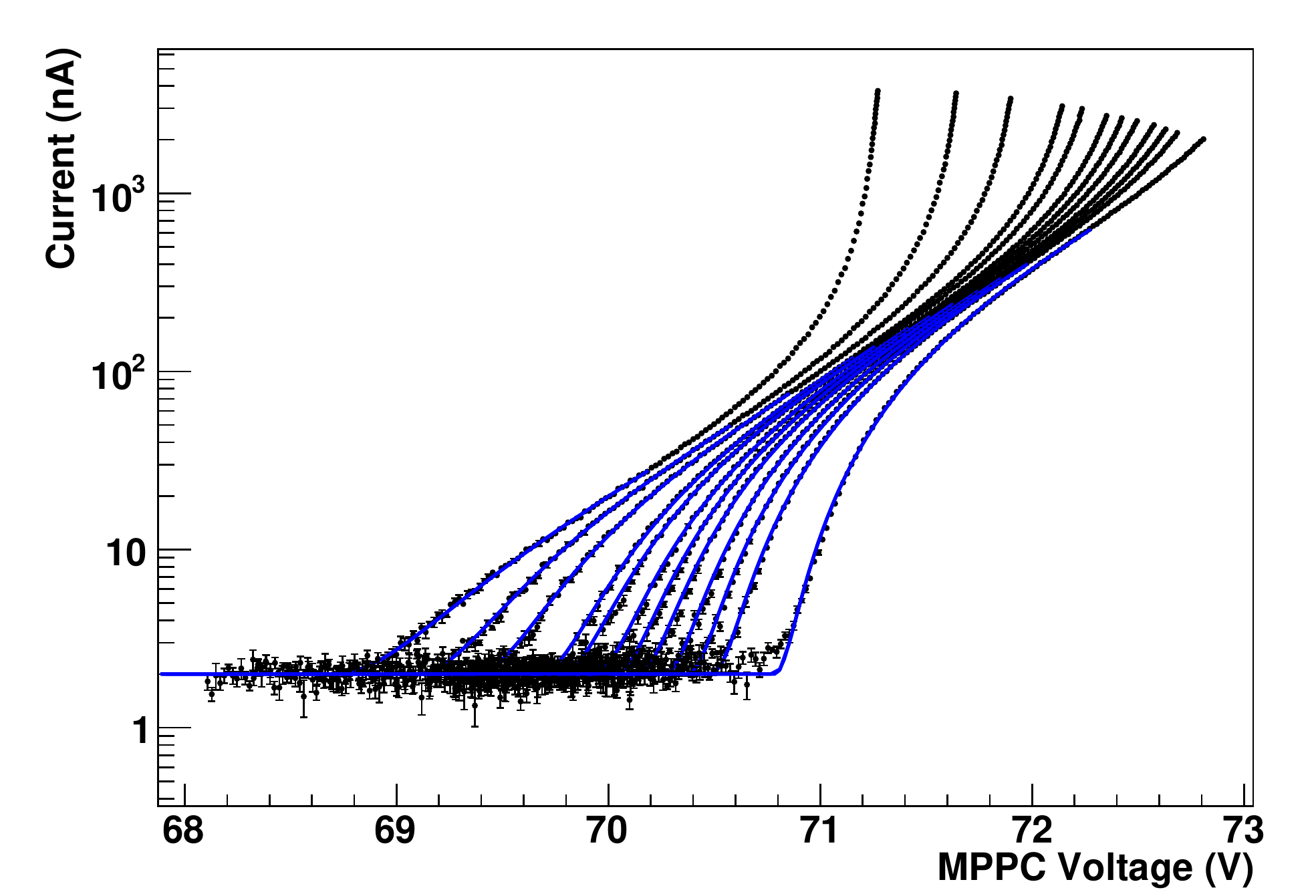}
  \caption{A comparison of I-V graphs and fits from a single MPPC (3x3 MPPC04) across a range of temperatures (from left to right: 1, 5, 10, 16, 17, 20, 21, 23, 25, 27, 29, and 34 $^\circ$C) using a logarithmic current scale. The blue curves are the best fit functions within their respective fit range.}
  \label{fig:logplot}
\vspace{8mm}
    \includegraphics[width=0.45\textwidth]{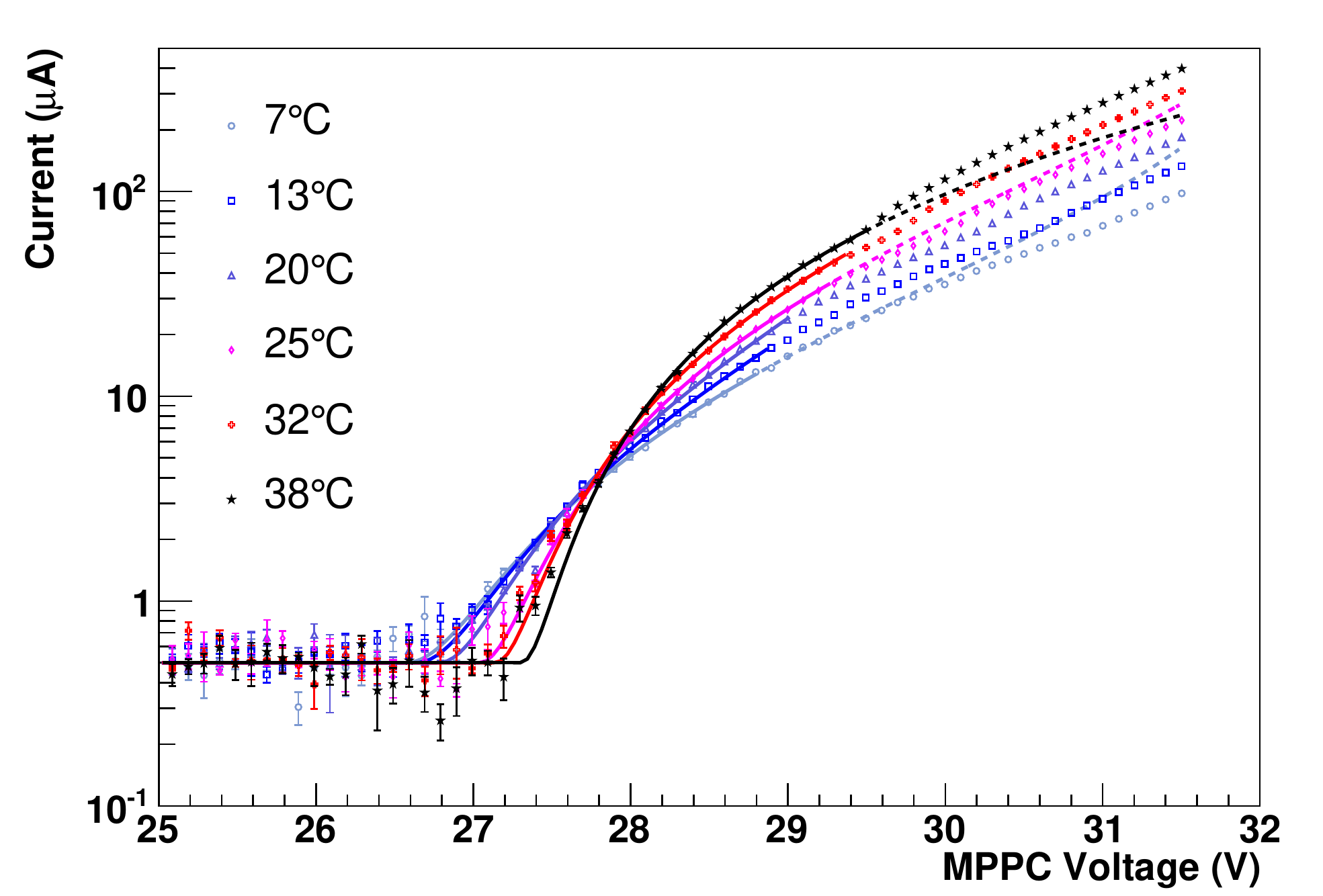}
  \caption{Current vs voltage at different temperatures for the SensL SPMArray. The fit functions are overlaid over the data.}
  \label{fig:SensL_IV}
\end{figure}

Figure~\ref{fig:T20_IV} shows the example of a good fit. The fits are very good above zero $^\circ$C. At sub-zero temperatures, however, the behaviour of the MPPCs was not well described by the above equation, and a reliable fit providing any useful information was difficult to obtain. This was particularly true for the I-V curves at -60$^\circ$C, where the region of the curve just above the breakdown voltage is a radically different shape than at higher temperatures as shown in Figure~\ref{fig:T-60_IV}. The T2K MPPC characterization paper did not explore MPPC properties below 0$^\circ$C, so this is not entirely unexpected \cite{T2Kmppc}. Since the operating temperature of the MPPCs will be well above 0$^\circ$C, this is of little concern. Figure\ref{fig:logplot} shows a the current dependence as a function of voltage for 12 different temperatures above 0$^\circ$C using a logarithmic scale. The curves shift to the right following the breakdown voltage and also become steeper with increasing temperature due to the increasing dark noise rate.

The SensL data behave somewhat differently as shown in Figure \ref{fig:SensL_IV}. The breakdown voltage increases smoothly with increasing temperature but the curve cross each other at about 28 V of applied bias. Nevertheless, the SensL data remain well modeled by the fit function up to 2.25V above $V_{bd}$, however the extensions of the function beyond this tend to be less accurate than those with the MPPCs. The fit function most closely matched the SensL data at temperatures around 28$^\circ$C.

Some data sets show a larger current than expected by the fit function in the vicinity of the breakdown voltage as is clearly visible in the right hand most data set in Figure~\ref{fig:logplot}. This deviation can be interpreted as the linear gain regime used prior to reaching the Geiger mode regime. It is not accounted for in the fit function because no satisfactory function has been derived. The main issue is that  while the gain dependence could be inferred from\cite{apd} accounting for the quenching resistor, the dark noise rate dependence below breakdown voltage is not known. This deviation is not expected to significantly affect the fit parameters.

\begin{table}
\centering
\caption{Characteristic values of the MPPCs determined from the fitted I-V curves}
\begin{tabular}{|c|c|c|c|}
\hline
\multirow{2}{*}{Sensor ID} & $[{V_{bd}}]_{20 ^\circ C}$ & \multirow{2}{*}{$[N_{ca}]_{20 ^\circ C}$} & $[DN\cdot C]_{20 ^\circ C}$ \\ 
& (V) & & (nA/$V^2$) \\ \hline
3x3 MPPC01 & 70.24 $\pm$ 0.01 & 0.122 $\pm$ 0.001 & 51.7 $\pm$ 0.3 \\ \hline
3x3 MPPC02 & 70.28 $\pm$ 0.01 & 0.125 $\pm$ 0.001 & 51.9 $\pm$ 0.2 \\ \hline
3x3 MPPC03 & 70.25 $\pm$ 0.01 & 0.127 $\pm$ 0.001 & 52.1 $\pm$ 0.2 \\ \hline
3x3 MPPC04 & 69.93 $\pm$ 0.01 & 0.125 $\pm$ 0.001 & 56.4 $\pm$ 0.2 \\ \hline
3x3 MPPC05 & 70.34 $\pm$ 0.01 & 0.113 $\pm$ 0.001 & 57.2 $\pm$ 0.2 \\ \hline
3x3 MPPC06 & 70.21 $\pm$ 0.01 & 0.119 $\pm$ 0.001 & 56.3 $\pm$ 0.3 \\ \hline
3x3 MPPC07 & 69.96 $\pm$ 0.01 & 0.118 $\pm$ 0.001 & 57.2 $\pm$ 0.4 \\ \hline
3x3 MPPC08 & 70.19 $\pm$ 0.01 & 0.121 $\pm$ 0.001 & 52.0 $\pm$ 1 \\ \hline
\hline
1x4 MPPC03 & 70.18 $\pm$ 0.01 & 0.105 $\pm$ 0.001 & 47.5 $\pm$ 0.4 \\ \hline
1x4 MPPC04 & 70.28 $\pm$ 0.01 & 0.109 $\pm$ 0.001 & 43.9 $\pm$ 0.5 \\ \hline
1x4 MPPC05 & 70.34 $\pm$ 0.01 & 0.105 $\pm$ 0.001 & 38.2 $\pm$ 0.4 \\ \hline
1x4 MPPC06 & 70.33 $\pm$ 0.01 & 0.110 $\pm$ 0.001 & 37.8 $\pm$ 0.5 \\ \hline
1x4 MPPC07 & 70.34 $\pm$ 0.01 & 0.111 $\pm$ 0.001 & 48.4 $\pm$ 0.6 \\ \hline
1x4 MPPC08 & 70.32 $\pm$ 0.01 & 0.103 $\pm$ 0.001 & 43.4 $\pm$ 1 \\ \hline
1x4 MPPC09 & 70.17 $\pm$ 0.01 & 0.103 $\pm$ 0.001 & 38.5 $\pm$ 0.9 \\ \hline
1x4 MPPC10 & 70.22 $\pm$ 0.01 & 0.108 $\pm$ 0.001 & 40.9 $\pm$ 0.6 \\ \hline
\hline
SensL SA1 & 27.03 $\pm$ 0.02 & 0.027 $\pm$ 0.004 & 4.7 10$^3$ $\pm$ 200 \\ \hline
SensL SA2 & 26.96 $\pm$ 0.01 & 0.032 $\pm$ 0.002 & 4.8 10$^3$ $\pm$ 100 \\ \hline
SensL SL1 & 27.09 $\pm$ 0.03 & 0.057 $\pm$ 0.003 & 4.1 10$^3$  $\pm$ 200 \\ \hline
SensL SL2 & 27.087 $\pm$ 0.005 & 0.0570 $\pm$ 0.0007 & 3.8 10$^3$  $\pm$ 300 \\ \hline
\end{tabular}
\label{tab:ParamTable1}
\end{table}
From these fits, the values of the parameters involved, $V_{bd}$, $N_{ca}$, $DN\cdot C$ and the constant $\beta$ were extracted, along with their uncertainties, as seen in Table\ref{tab:ParamTable1}. The dark noise rate is in qualitative agreement with the measure provided by Hamamatsu for each MPPC at 25$^\circ$C and 7.5 $10^5$ gain assuming 90fF for the pixel capacitance. The rate of correlated avalanches is in good qualitative agreement with \cite{T2Kmppc} when summing the contribution of cross-talk and after-pulsing. This technique could also be used to measure photo-detection efficiency with a source bright enough to yield a photo-current comparable to the dark current. The fit technique allows the subtraction of the correlated avalanche contribution but it is important to ensure that the PPD does not saturate.

%
%
\begin{figure} [t]
  \centering
    \includegraphics[width=0.45\textwidth]{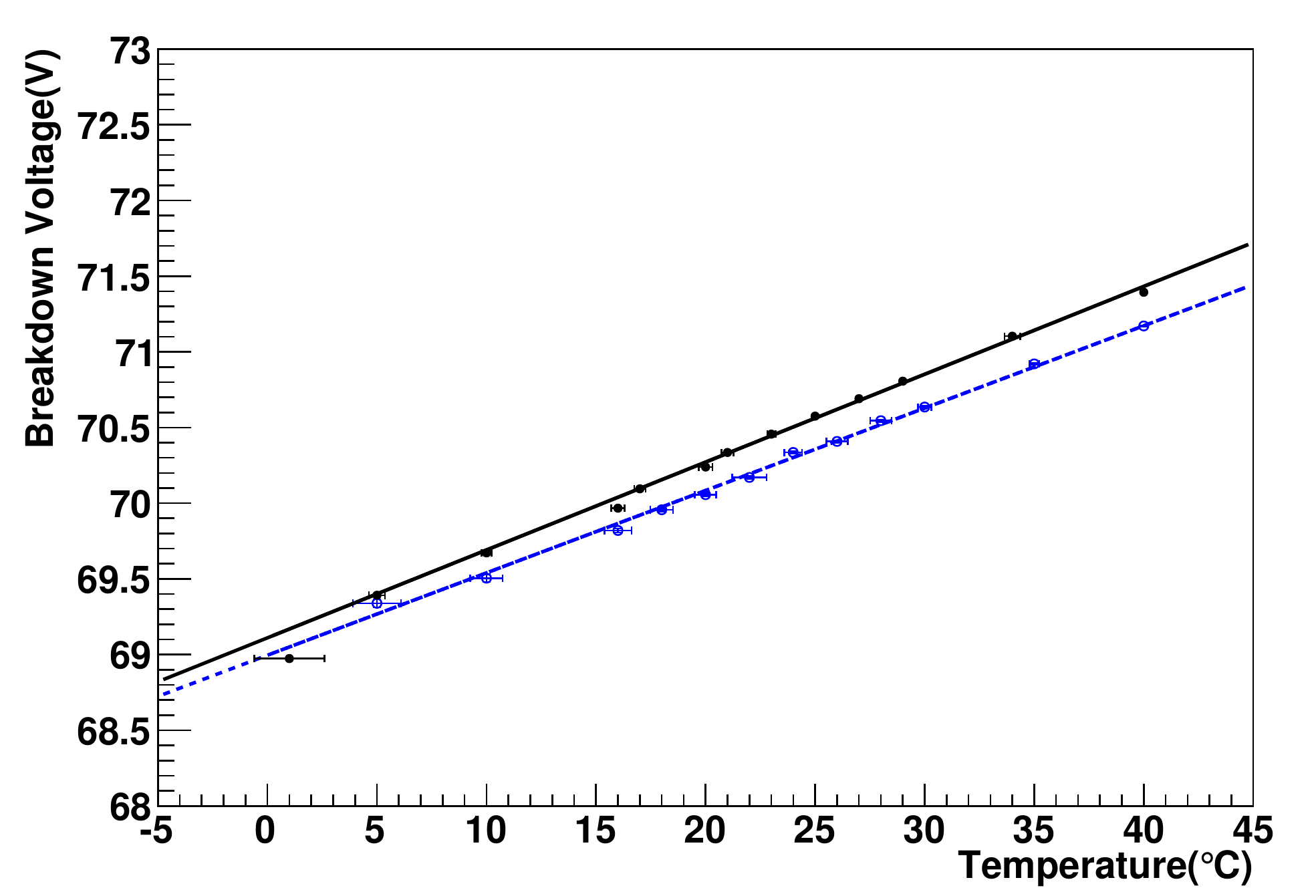}
  \caption{Breakdown voltage versus temperature for 3x3 mm$^2$ MPPC 01 (filled circles, solid line) and 1x4 mm$^2$ MPPC 04 (open circles, dashed line).}
  \label{fig:Vbd_T_plot}
\end{figure}
\section{Determining Temperature Dependence of MPPC Characteristics}

\begin{table}
\centering
\caption{Results of fits of temperature dependence for $N_{ca}$ and $V_{bd}$ for all MPPCs. $A\cdot C$ was scaled per unit area in order to compare different devices.}
\begin{tabular}{|c|c|c|c|c|}
\hline
\multirow{2}{*}{Sensor ID} & $\Delta V_{bd}$ & $\Delta N_{ca}$ & $A\cdot C$ & \multirow{2}{*}{E (eV)}\\ 
 & (mV/$^\circ$C) & $10^{-3}/^\circ C$ & (nA/$V^2$/mm$^{2}$) & \\ \hline
3x3 MPPC01 & 58.0 $\pm$ 0.4 & -0.83 $\pm$ 0.03 & 9.3 $\pm$ 0.08 & 1.32 $\pm$ 0.02 \\ \hline
3x3 MPPC02 & 58.2 $\pm$ 0.4 & -1.06 $\pm$ 0.03 & 9.4 $\pm$ 0.08 & 1.32 $\pm$ 0.02 \\ \hline
3x3 MPPC03 & 58.4 $\pm$ 0.4 & -1.01 $\pm$ 0.04 & 9.7 $\pm$ 0.09 & 1.34 $\pm$ 0.02 \\ \hline
3x3 MPPC04 & 57.6 $\pm$ 0.4 & -1.04 $\pm$ 0.03 & 10.2 $\pm$ 0.09 & 1.32 $\pm$ 0.02 \\ \hline
3x3 MPPC05 & 59.0 $\pm$ 0.4 & -1.30 $\pm$ 0.04 & 10.1 $\pm$ 0.09 & 1.34 $\pm$ 0.02 \\ \hline
3x3 MPPC06 & 58.3 $\pm$ 0.5 & -1.07 $\pm$ 0.04 & 10.3 $\pm$ 0.09 & 1.32 $\pm$ 0.02 \\ \hline
3x3 MPPC07 & 58.3 $\pm$ 0.4 & -1.39 $\pm$ 0.04 & 10.6 $\pm$ 0.09 & 1.38 $\pm$ 0.02 \\ \hline
3x3 MPPC08 & 58.5 $\pm$ 0.5 & -0.57 $\pm$ 0.07 & 9.7 $\pm$ 0.1 & 1.37 $\pm$ 0.02 \\ \hline
\hline
1x4 MPPC03 & 53.4 $\pm$ 0.6 & -0.28 $\pm$ 0.03 & 12.9 $\pm$ 0.18 & 1.18 $\pm$ 0.02 \\ \hline
1x4 MPPC04 & 54.4 $\pm$ 0.7 & -0.06 $\pm$ 0.02 & 14.2 $\pm$ 0.2 & 1.16 $\pm$ 0.02 \\ \hline
1x4 MPPC05 & 53.7 $\pm$ 0.6 & -0.08 $\pm$ 0.04 & 10.5 $\pm$ 0.15 & 1.20 $\pm$ 0.02 \\ \hline
1x4 MPPC06 & 53.8 $\pm$ 0.6 & -0.37 $\pm$ 0.02 & 11.0 $\pm$ 0.15 & 1.22 $\pm$ 0.02 \\ \hline
1x4 MPPC07 & 53.2 $\pm$ 0.7 & -0.93 $\pm$ 0.02 & 12.7 $\pm$ 0.18 & 1.18 $\pm$ 0.02 \\ \hline
1x4 MPPC08 & 54.9 $\pm$ 0.6 & -0.18 $\pm$ 0.02 & 14.0 $\pm$ 0.2 & 1.21 $\pm$ 0.02 \\ \hline
1x4 MPPC09 & 55.6 $\pm$ 0.9 & -0.20 $\pm$ 0.03 & 10.8 $\pm$ 0.15 & 1.24 $\pm$ 0.02 \\ \hline
1x4 MPPC10 & 57 $\pm$ 1 & -0.08 $\pm$ 0.03 & 11.5 $\pm$ 0.18 & 1.24 $\pm$ 0.02 \\ \hline
\hline
SPMArray4 1  & 20.0 $\pm$ 0.4 & -1.2 $\pm$ 0.1 & 32.2  $\pm$ 0.8 & 0.79 $\pm$ 0.01 \\ \hline
SPMArray4 2 & 21.8 $\pm$ 0.5 & -1.7 $\pm$ 0.1 & 33.1  $\pm$ 0.8 & 0.84 $\pm$ 0.01 \\ \hline
ArraySL-4 1 & 14.2 $\pm$ 0.7 & -2.4 $\pm$ 0.2 & 26.0  $\pm$ 0.7 & 0.73 $\pm$ 0.02 \\ \hline
ArraySL-4 2 & 16.2 $\pm$ 0.2 & 2.1 $\pm$ 0.2 & 28.5 $\pm$ 0.7 & 0.85 $\pm$ 0.01 \\ \hline
\end{tabular}
\label{tab:ParamTable2}
\end{table}

A full temperature scan was performed to determine the temperature dependence of the breakdown voltage, dark noise rate, and average number of correlated avalanches. The manner in which the shape of an I-V curve and its breakdown voltage vary with temperature for a given MPPC is shown in Figure~\ref{fig:logplot}. The curves clearly change shapes between 1 and 34 $^\circ$C becoming steeper with increasing temperature close to the breakdown voltage due to increasing dark noise rate, but flattening at higher over-voltage due to the decreasing rate of correlated avalanches. The fit parameters shown in table\ref{tab:ParamTable2} demonstrate this interpretation quantitatively. The same conclusion can be made for the SensL's devices.

\begin{figure} [t]
  \centering
  \includegraphics[width=0.45\textwidth]{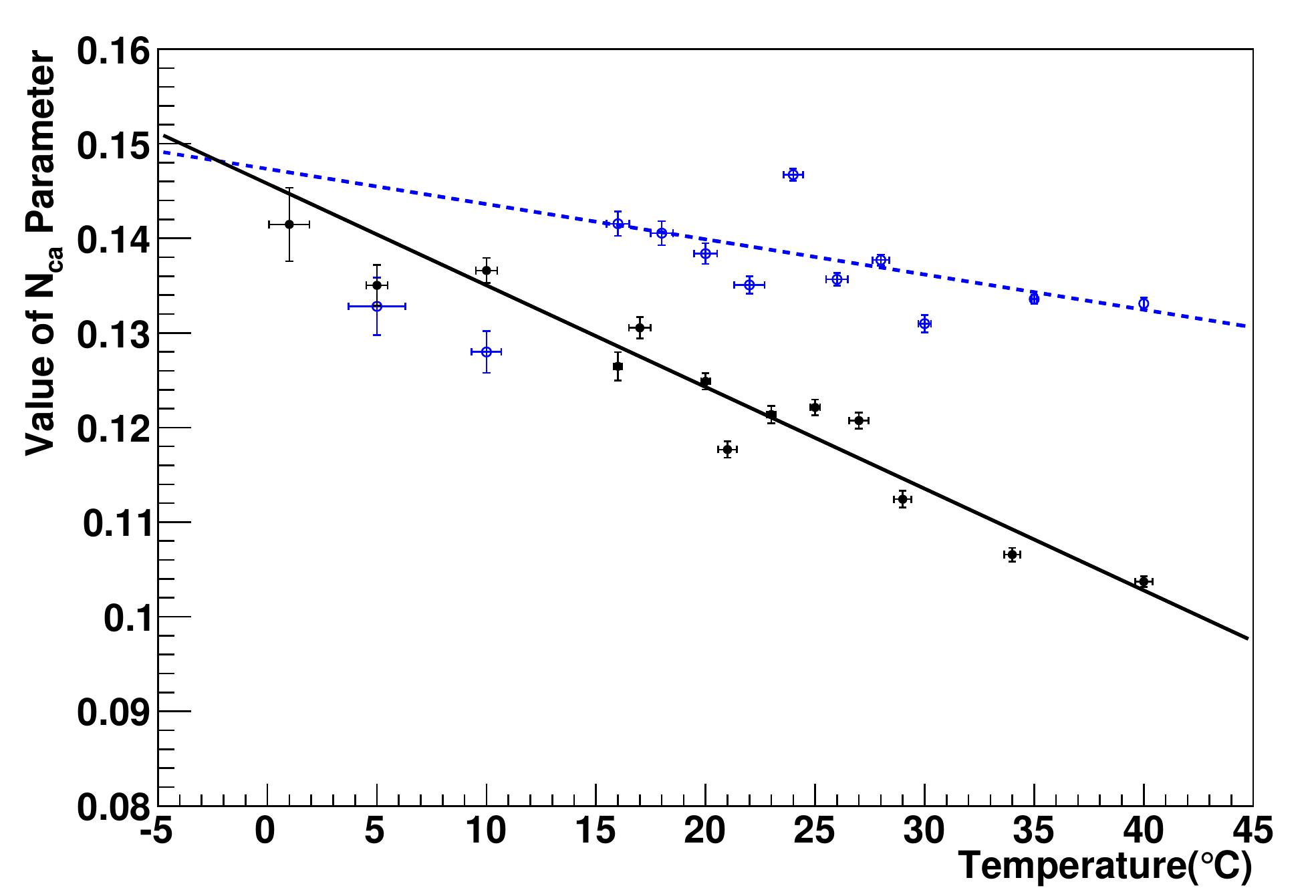}
  \caption{Average number of correlated avalanches per parent avalance ($N_{ca}$ parameter) versus temperature for 3x3 mm$^2$ MPPC 02 (filled circles, solid line) and 1x4 mm$^2$ MPPC 06 (open circles, dashed line).}
  \label{fig:Nca_T_plot}
\vspace{8mm}
   \includegraphics[width=0.45\textwidth]{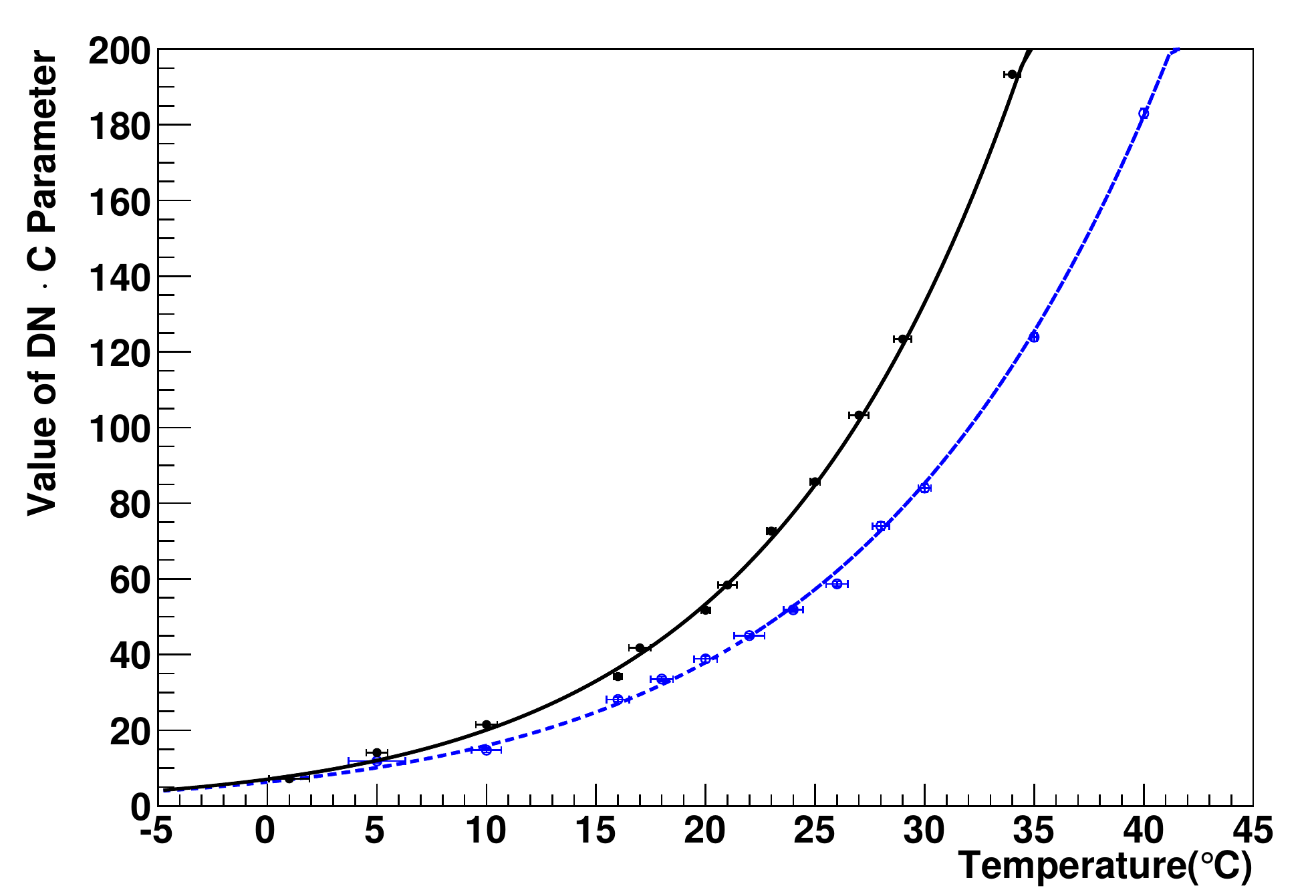}
  \caption{Product of the dark noise rate and pixel capacitance nA/V$^2$ versus temperature for 3x3 mm$^2$ MPPC 01 (filled circles, solid line) and 1x4 mm$^2$ MPPC 04 (open circles, dashed line).}
\label{fig:DNC_T_plot}
\end{figure}

 Figure~\ref{fig:Vbd_T_plot} shows the dependence of $V_{bd}$ with temperature for one MPPC of each size. $V_{bd}$ increased by around 55 mV/$^\circ$C for the 1x4 mm$^2$ MPPCs, and around 58 mV/$^\circ$C for the 3x3 mm$^2$ MPPCs, closely following a linear distribution. This falls within the range of measurements made by other collaborations. Figure~\ref{fig:Nca_T_plot} shows the dependence of the average number of correlated avalanche parameter $N_{ca}$ for one  MPPC of each size. $N_{ca}$ shows a decreasing trend with increasing temperature for all MPPCs and most SensL SPMs. This trend was found to be approximately linear, as was observed in the previous MPPC characterization paper \cite{T2Kmppc}. It was also observed that the $N_{ca}$ value decreases more rapidly with increasing temperature for the 3x3 mm$^2$ MPPCs than for the 1x4 mm$^2$ MPPCs. Due to the nature of the parameter's role in the current fitting function, the fit is quite sensitive to its value, and so it has a high variance. The dependence of the product of the dark noise rate and pixel capacitance with temperature is shown in Figure~\ref{fig:DNC_T_plot}. It was fitted using the relationship introduced in~\cite{T2Kmppc}:
\begin{equation}
    DN(T)\cdot C = C\cdot A\cdot \Big(\frac{T}{298}\Big)^{\frac{3}{2}} \cdot e^{-(\frac{E}{2kT} - \frac{E}{2k\cdot 298})}
\end{equation}
the band gap energy $E$ of the MPPCs was measured, along with the value of the product of the leading constant $A$ with the pixel capacitance. While there is some variation of the value of $E$ within MPPCs of the same type, there seems to be a significant difference between types. The average measurements were 1.20 eV for the 1x4 mm$^2$ MPPCs and 1.34 eV for the 3x3 mm$^2$ MPPCs. These are similar to the value of 1.136 eV presented in the T2K characterization paper, in which a 1.3x1.3 mm$^2$ MPPC was used \cite{T2Kmppc}. The product of the leading constant $A$ and the pixel capacitance was measured to average 48.7 nA/$V^2$ for the 1x4 mm$^2$ MPPCs and 89.2 nA/$V^2$ for the 3x3 mm$^2$ MPPCs, on average. Given that the capacitance of each pixel is known to have a value of 90fF \cite{T2KExp} \cite{mppcSpec}, the average values of the constant A can be calculated as 541 kHz/V for the 1x4 mm$^2$ MPPCs and 991 kHz/V for the 3x3 mm$^2$ MPPCs. The dark noise rate shows an approximate scaling with MPPC area, the 3x3 mm$^2$ devices having a smaller rate per mm$^2$ because they were selected for their low dark noise rate within a batch of 30 devices.

\begin{figure}[t]
  \centering
    \includegraphics[width=0.45\textwidth]{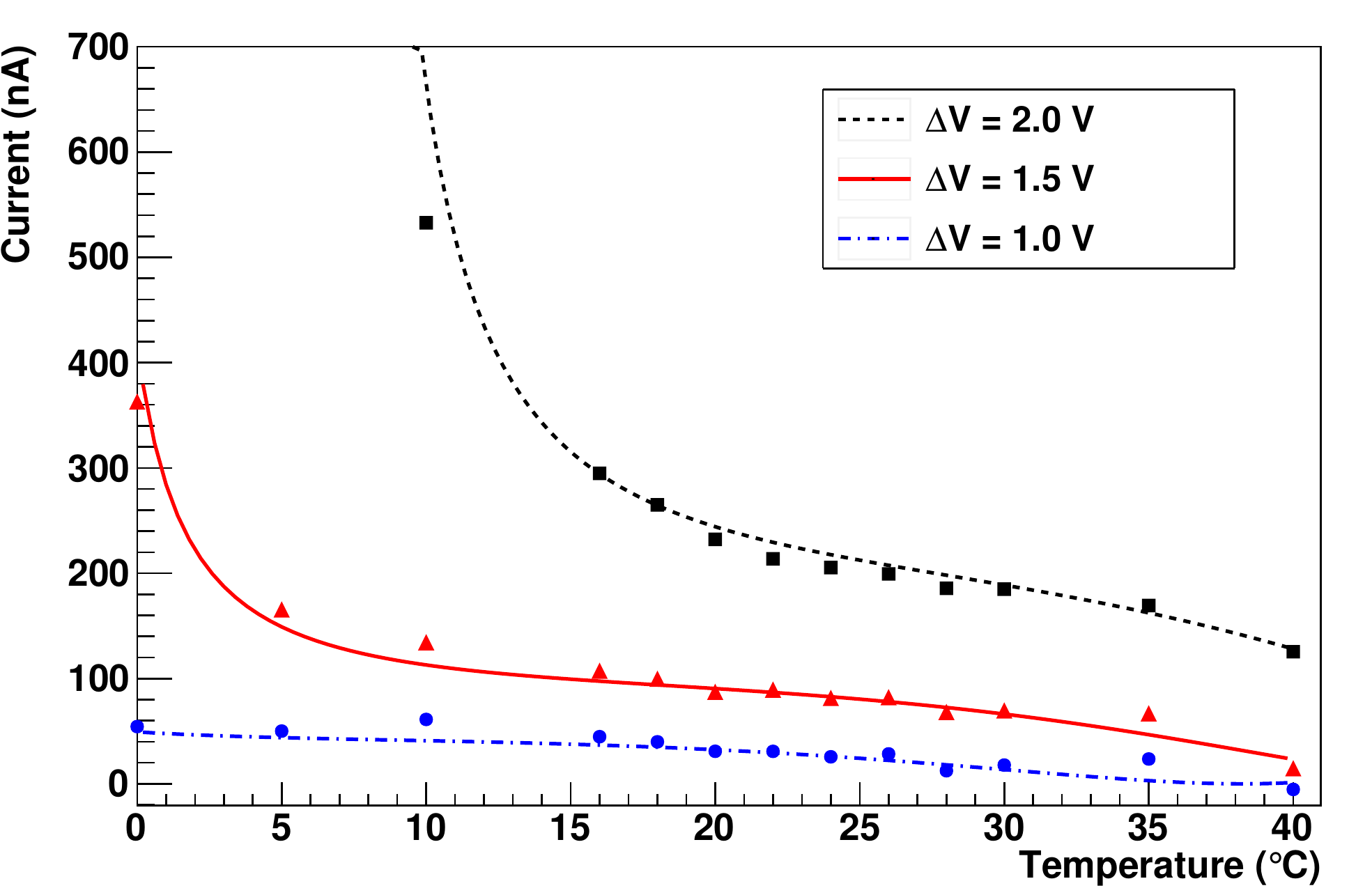}
  \caption{Current vs temperature curves generated for 1x4 mm$^2$ MPPC 05 at 20 $^\circ$C. The curves were calculated using previously extracted fit parameters. The plotted points represent measurements extracted from the I-V curves at each temperature.}
  \label{fig:I-T_example}
\end{figure}

For stable operation of the PPD, it is important to operate at constant over-voltage, which requires either to maintain the temperature constant or to
adjust the operating voltage should the temperature vary by more than a few $^\circ$C. In either case, a temperature probe is required. In the absence of significant photo-current compared to the dark current, one could conceive using the PPD current measure to infer temperature changes. To investigate this solution, the previous form for current as a function of overvoltage can be rewritten to be a function of temperature, by replacing each of the temperature-dependent constants with the expressions determined through fits. The operating voltage is kept constant when measuring temperature but it could be subsequently updated should a significant temperature change be detected. The value of the initial operating voltage is determined by the starting temperature (which defines the initial breakdown voltage) and the desired overvoltage. The resulting curve shows how the MPPC dark current will vary with temperature if no adjustments to voltage are made. Unfortunately, the current change is small as shown in Figure~\ref{fig:I-T_example} because the increase in dark noise rate with increasing temperature is compensated by the over-voltage decrease. The change in current that results from a 1$^\circ$C  to around 20 $^\circ$C is about 3 nA. The current readout of the control card used in this paper was found to have a standard deviation of 3.9 nA , which would require 16 measurements to be accurate within 1 nA. A resolution of 0.7nA was achieved with a newer board, which would require only 4 to 5 current measurements to achieve the desired accuracy but that level of precision may be hard to achieve when using a large number of devices. Hence, we conclude that the MPPC current measure is not sensitive enough to easily probe temperature variations when keeping the operating voltage constant. It is possible to measure unambiguously the MPPC temperature by performing a scan of the operating voltage with prior knowledge of the fit parameters (through a calibration procedure), but such a scan would introduce dead time in most experiments. On the other hand, this lack of temperature sensitivity may prove advantageous when using MPPCs to monitor the average power of light sources bright enough to yield a photo-current comparable to the dark current. The same conclusion can be drawn for the SensL devices, the sensitivity to temperature being even smaller as the current vs voltage overlaps across a wide range of temperatures.

\section{Conclusion}

It has been found that a voltage scan of the dark current, along with the fitting of the determined formula for current as a function of overvoltage, is a very effective tool for determining the breakdown voltage, dark noise rate constant, and the average number of correlated avalanches. The photo-detection efficiency could also be determined in the same way using an appropriate light source. The values and temperature dependences of the Hamamatsu MPPC and SensL SPM parameters have been shown to be in good agreement with previous measurements done measuring individual pulses, which is significantly more complex. Hence, this paper introduces a simple way to characterize PPDs using cost effective DC current sensitive electronics. 

This paper also shows that a method of monitoring temperature variations of MPPCs and SensL SPM exclusively through dark current measurement is difficult to implement because it requires sub-nA precision, which is hard to achieve for a large number of channels. On the other hand, the lack of temperature dependence allows the monitoring of the average power of bright light sources without requiring a significant temperature compensation scheme within $\pm 5 ^\circ C$.






\bibliographystyle{IEEEtran}

\clearpage

\end{document}